%% file: Prototype_Paper.tex
\documentclass[submission,copyright,creativecommons]{eptcs}
\providecommand{\event}{AMMSE 2011}
\usepackage{breakurl}
\usepackage{graphicx}
\usepackage{caption}
\usepackage{subcaption}
\usepackage{color}
\usepackage{bold-extra}
\usepackage{listings, float}

\input{macros}

\newcommand{\mycolorcmd}[1]{\textcolor{black}{#1}}
\newcommand{\mcc}[1]{\mycolorcmd{#1}}

\newfloat{listing}{hbt}{lst}

\title{Prototyping the Semantics of a DSL using ASF+SDF:\\ 
       Link to Formal Verification of DSL Models}
  \author{Suzana Andova \qquad Mark van den Brand \qquad Luc Engelen
  \institute{
    Eindhoven University of Technology\\
    P.O. Box 513, 5600 MB, Eindhoven, The Netherlands\\
  }
  \email{\{S.Andova | M.G.J.v.d.Brand | L.J.P.Engelen\}@tue.nl}
}
\def\titlerunning{Prototyping the Semantics of a DSL using the ASF+SDF Meta-Environment}
\def\authorrunning{S. Andova, M.G.J. van den Brand \& L.J.P. Engelen}

\begin{document}
\maketitle

\begin{abstract}
 \input{Abstract}
\end{abstract}

\input{Introduction}

\input{DSL1}

\input{Prototyping_Semantics1}

\input{Verification}

\input{Related_Work}

\input{Conclusions_and_Future_Work}

\bibliographystyle{eptcs}

\bibliography{Prototype_Paper}

\end{document}

%% file: macros.tex
\mathcode`:="003A       
\mathcode`;="203B       
\mathcode`?="003F       
\mathcode`|="026A       
\mathcode`<="4268       
\mathcode`>="5269       

\mathchardef\ls="213C   
\mathchardef\gr="213E   
\mathchardef\uparrow="0222      
\mathchardef\downarrow="0223    

\newcommand{\nop}[1]{}

\newcommand{\SLCO}{\mbox{\tt SLCO}}
\newcommand{\LTS}{\mbox{\tt LTS}}
\newcommand{\DOT}{\mbox{\tt Dot}}
\newcommand{\CSs}{\mbox{\tt CS}}
\newcommand{\CS}{\mbox{\tt CS}}

\newcommand{\Initial}{\mbox{\tt Initial}}
\newcommand{\State}{\mbox{\tt State}}

\newcommand{\rec}{\mbox{\tt receive}}

\newcommand{\Stop}{\mbox{\tt Stop}}
\newcommand{\fromport}{\mbox{\tt from}}

\newcommand{\op}{\mbox{\it p}}
\newcommand{\oq}{\mbox{\it q}}

%% file: Abstract.tex
\mcc{A formal definition of the semantics of a domain-specific language (DSL) is a key prerequisite for the verification of the correctness of models specified using such a DSL and of transformations applied to these models.
For this reason, we implemented a prototype of the semantics of a DSL for the specification of systems consisting of concurrent, communicating objects.
Using this prototype, models specified in the DSL can be transformed to labeled transition systems (LTS).}
\mcc{This approach of transforming models to LTSs allows us to apply existing tools for visualization and verification to models with little or no further effort.}
The prototype is implemented using the ASF+SDF Meta-Environment, an IDE for the algebraic specification language ASF+SDF, which offers efficient execution of the transformation as well as the ability to read models and produce LTSs without any additional pre or post processing. 

%% file: Introduction.tex
\section{Introduction}
\label{sec:Introduction}

Domain-specific languages (DSL) and model transformations are the key concepts in model driven engineering~\cite{Schmidt2006}.
A DSL is a language that offers, through appropriate notations and abstractions, expressive power focused on, and usually restricted to, a particular problem domain~\cite{Deursen2000}.
A DSL enables domain experts to develop models using concepts in their own domain, rather than concepts provided by existing formalisms, which typically do not provide the required or correct abstractions.
DSL models can be further transformed into models in other languages using model transformations.
Model transformations can, for example, be used to transform DSL models into different implementations, each serving different purposes, such as validation, execution, testing, and visualization.

\mcc{In our previous work, we have designed the Simple Language of Communicating Objects ($\SLCO$)~\cite{Amstel2010_slco}, which provides constructs for specifying systems consisting of objects that operate in parallel and communicate with each other.}
The structure of a system is modeled using classes and their behavior is modeled by state machines.
Simultaneously to the development of the language, we implemented a number of model transformations to other languages as well as within the $\SLCO$ language.
\mcc{The goal of the previous work was to generate, for a given $\SLCO$ model, (i)~its implementation in NQC for execution on the Lego Mindstorms platform~\cite{Baum2003}, (ii)~its implementation in POOSL~\cite{Theelen2007} for model simulation, and (iii)~its implementation in Promela~\cite{DBLP:journals/tse/Holzmann97} for formal verification by means of the SPIN model checker~\cite{DBLP:journals/tse/Holzmann97}.}
All three of these languages have semantic properties different from the semantics of $\SLCO$, and therefore, several semantic gaps needed to be bridged~\cite{Amstel2008}.
These semantic gaps are bridged using model transformations, that transform $\SLCO$ models into $\SLCO$ models with equivalent observable behavior.

\mcc{The transformations from $\SLCO$ to NQC, POOSL, and Promela provide a partial transformational description of the semantics of $\SLCO$, because each of these transformations only deals with a subset of $\SLCO$.
One of the goals of the work presented in this paper is to define the operational semantics of the entire language.}

As for model transformations within $\SLCO$, one way to reason about their correctness is to compare or relate the $\SLCO$ models before and after the transformation, by comparing or relating their implementations.
If, in addition, the implementation language is supported by a toolset allowing for model reduction, rather larger $\SLCO$ models can be handled, either for analysis or for model comparison.
As Promela and SPIN do not provide support for these features, we have been motivated to look for an alternative solution, which would impose no restrictions on the $\SLCO$ expressiveness, too.

\begin{figure}[hbt]
\centering
\includegraphics[width=.5\columnwidth]{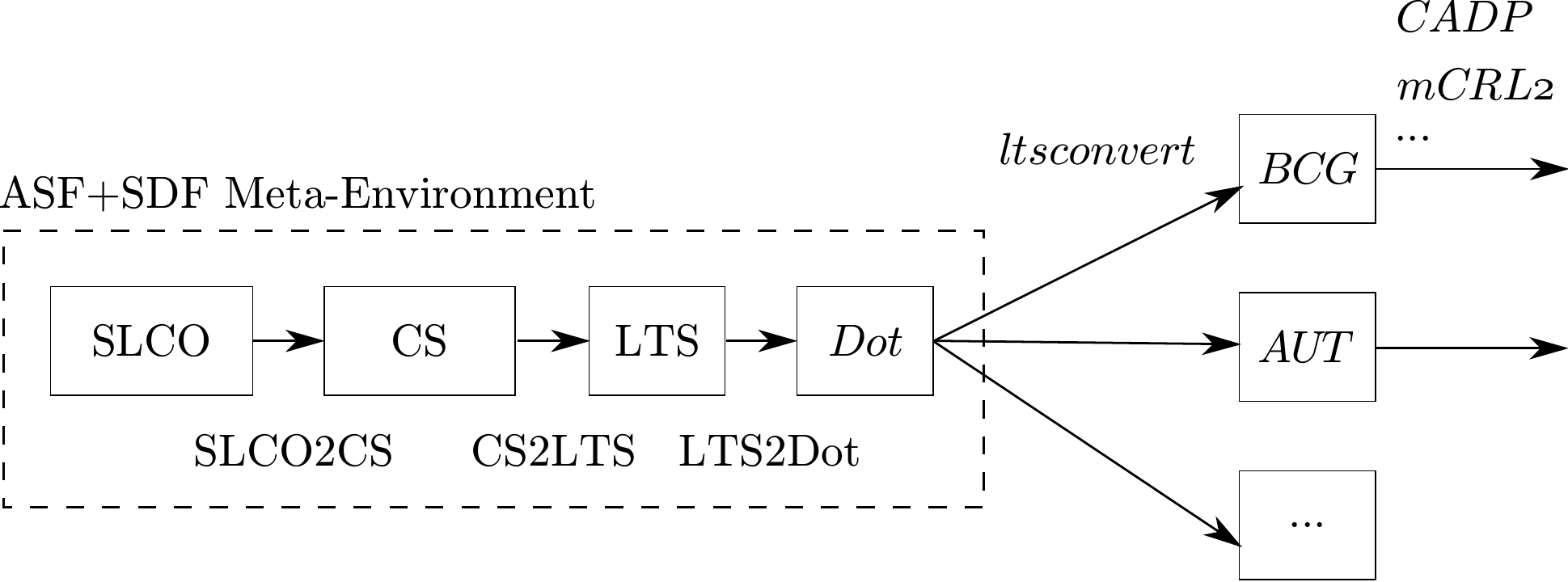}
\caption{Overview of languages and tools}
\label{fig:languages_overview}
\end{figure}

In this paper we link the $\SLCO$ language to a simple language for the representation of labeled transition systems, named simply $\LTS$.
Instead of selecting yet another target language supported by a toolset to which $\SLCO$ models will be transformed, we decided, first, to transform $\SLCO$ into $\LTS$ and, then, to link $\LTS$ to existing languages and toolsets (see Figure~\ref{fig:languages_overview}).
The reason is that $\LTS$, due to its simplicity, is much easier to link to other languages: languages that we have selected as appropriate for our purposes, such as those described in Section~\ref{sec:verification_and_visualization}, but also other languages which may show useful in the future.
\mcc{In our case, we achieve this by a rather easy transformation from $\LTS$ to $\DOT$, a language for the graphical representation of graphs that is also used by third-party tools to represent LTSs.}
Thus, our main effort is on translating $\SLCO$ to $\LTS$.
Once the translation into $\LTS$ has been achieved, different tools supporting LTSs are also within reach for manipulation, visualization, and verification of $\SLCO$ models.
Our language transformations are defined and implemented in the ASF+SDF Meta-Environment~\cite{Brand:2001:ASF}.
We find its ability to generate command-line tools that can efficiently execute transformations as well as its ability to parse arbitrary context-free languages very beneficial and advantageous for our approach.
As shown in Figure~\ref{fig:languages_overview}, our $\SLCO$ environment at the back-end can be connected to various already developed toolsets for various purposes.

Yet another motivation to take this approach is the following: the implementation of the transformations from $\SLCO$ models to LTSs gives a prototype of the semantics of $\SLCO$.
As shown in Figure~\ref{fig:languages_overview}, there is an intermediate representation language used, called $\CSs$, and therefore, two main transformation steps: $\SLCO 2 \CS$ and $\CS 2 \LTS$.
While objects in an $\SLCO$ model are defined separately and communicate via channels, in the $\CSs$ representation of the same system the objects are merged into one (big) component, according to the communication as defined in the $\SLCO$ model.
Thus, $\CSs$ links, in a rather natural way, the two languages $\SLCO$ and $\LTS$.
The transformation from $\SLCO$ to $\CSs$ essentially forms the core of the prototype semantics of $\SLCO$, as the transformation from $\CSs$ to $\LTS$ is rather straightforward.
The implementation of this transformation has helped to gain a better understanding of the effect of various design decisions on the semantics of $\SLCO$, in a larger part by the verboseness of the intermediate language $\CSs$ and the visual representation of LTSs produced by external tools.
Based on the conditional rewrite rules which form the core of this transformation, the formal operational semantics of the language can be defined, but this goes beyond the scope of this paper.
While our environment is built around $\SLCO$, we believe that the same methodology can be applied to other DSLs.

\emph{Structure of the paper}
In Section~\ref{sec:dsl}, $\SLCO$ is briefly described.
In Section~\ref{sec:prototyping_semantics}, the main ingredients of the transformation environment, the languages and the tools, are introduced, and in Section~\ref{sec:verification_and_visualization}, we link this environment to existing languages and tools.
Section~\ref{sec:Related_Work} addresses the related work, and Section~\ref{sec:Conclusions_and_Future_Work} concludes the paper and discusses future work.

%% file: DSL1.tex
\section{The Simple Language of Communicating Objects - SLCO}
\label{sec:dsl}
\begin{figure}[hbt]
 \centering
 \includegraphics[width=0.7\columnwidth]{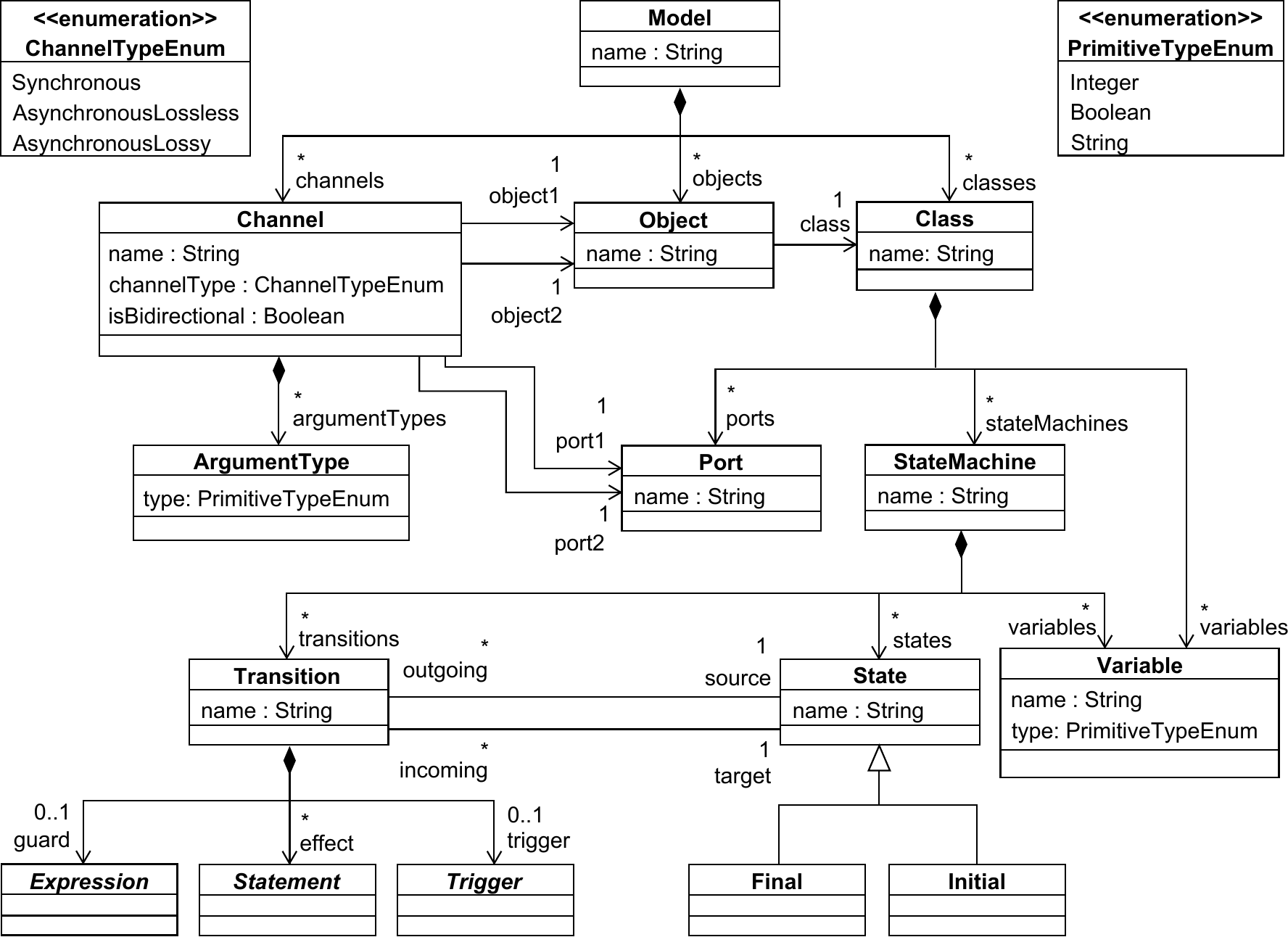}
 \caption{Overview of the $\SLCO$ metamodel}
 \label{fig:slco_mm}
\end{figure}

\begin{figure}[hbt]
\centering
 \begin{minipage}[b]{0.95\columnwidth}
  \centering
  \includegraphics[width=.7\columnwidth]{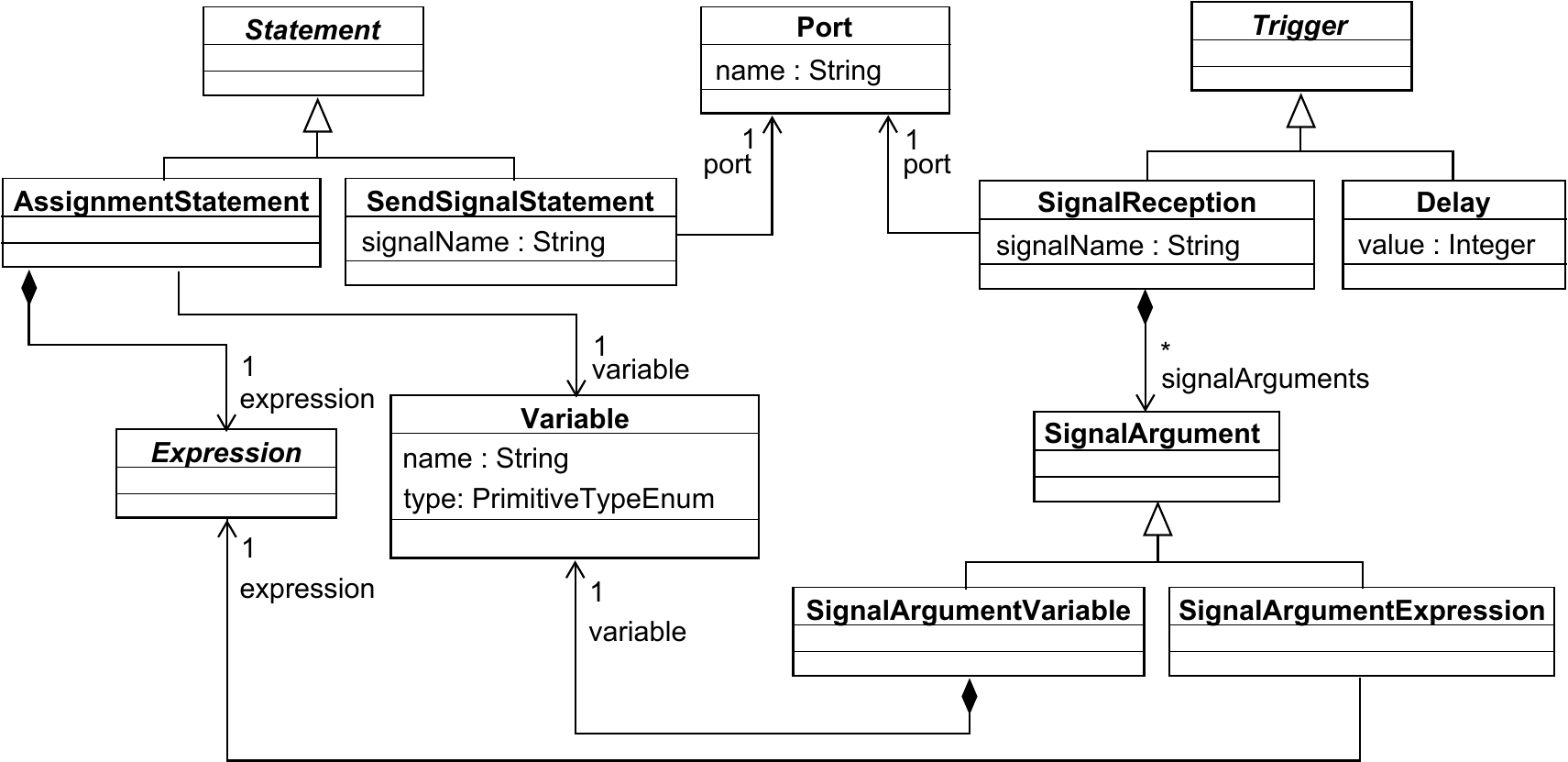}
  \subcaption{Statements and triggers}
  \label{fig:slco_stat_trig}
 \end{minipage}
 \begin{minipage}[b]{0.95\columnwidth}
  \centering
  \includegraphics[width=.7\columnwidth]{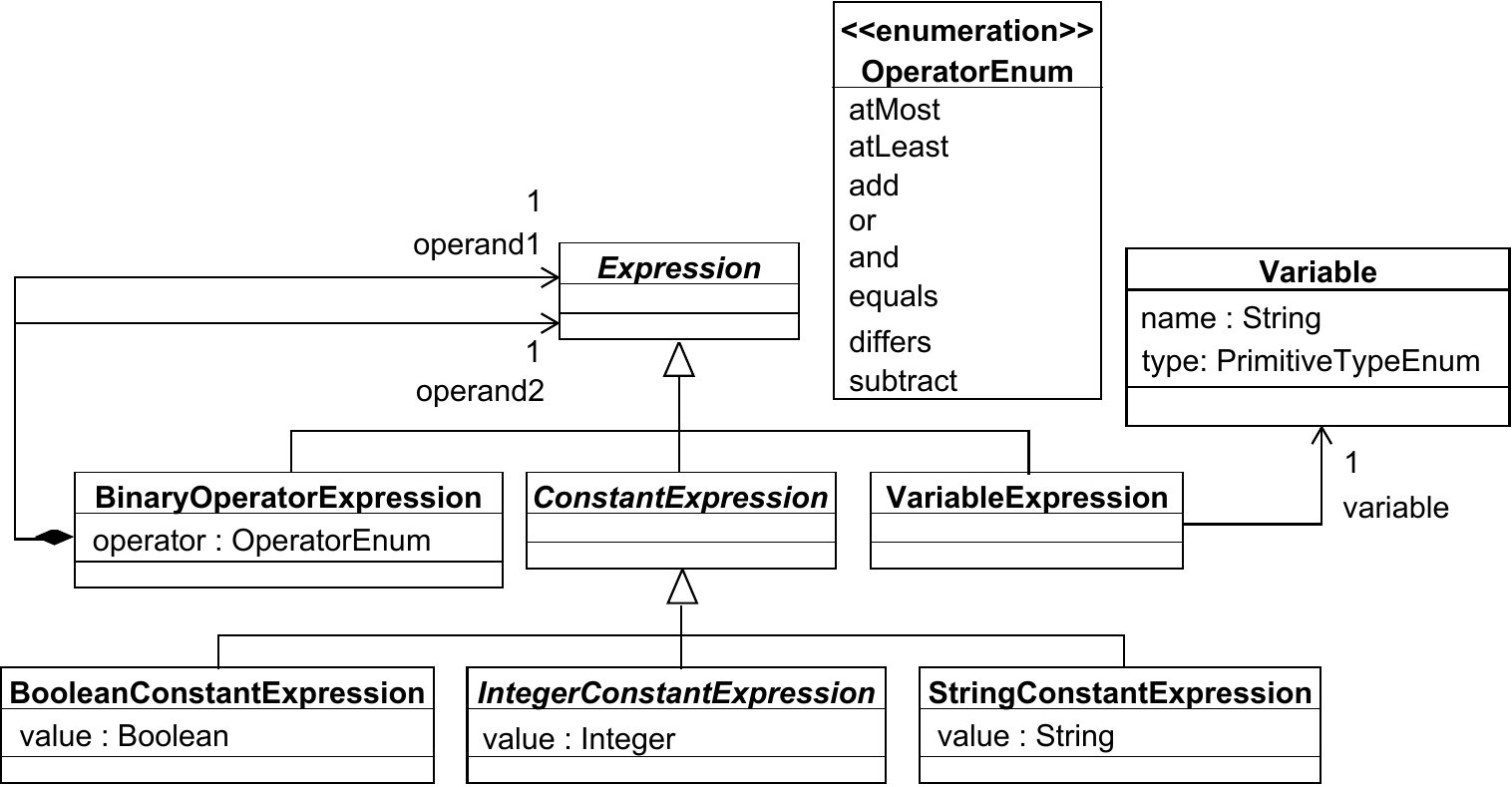}
  \subcaption{Expressions}
  \label{fig:slco_exp}
 \end{minipage}
 \caption{Statements, triggers, and expressions in $\SLCO$}
 \label{fig:slco_details}
\end{figure}

The \emph{Simple Language of Communicating Objects} ($\SLCO$) provides constructs for specifying systems consisting of objects that operate in parallel and communicate with each other.
Figure~\ref{fig:slco_mm} shows the main metaclasses and relations of the $\SLCO$ metamodel.
The remainder of the metamodel is shown in Figure~\ref{fig:slco_details}, which shows the subclasses of the abstract metaclasses \emph{Statement}, \emph{Trigger}, and \emph{Expression}, and the related metaclasses and relations.

\begin{figure}[hbt]
  \begin{minipage}[b]{.45\textwidth}
    \centering
\includegraphics[width=.6\columnwidth]{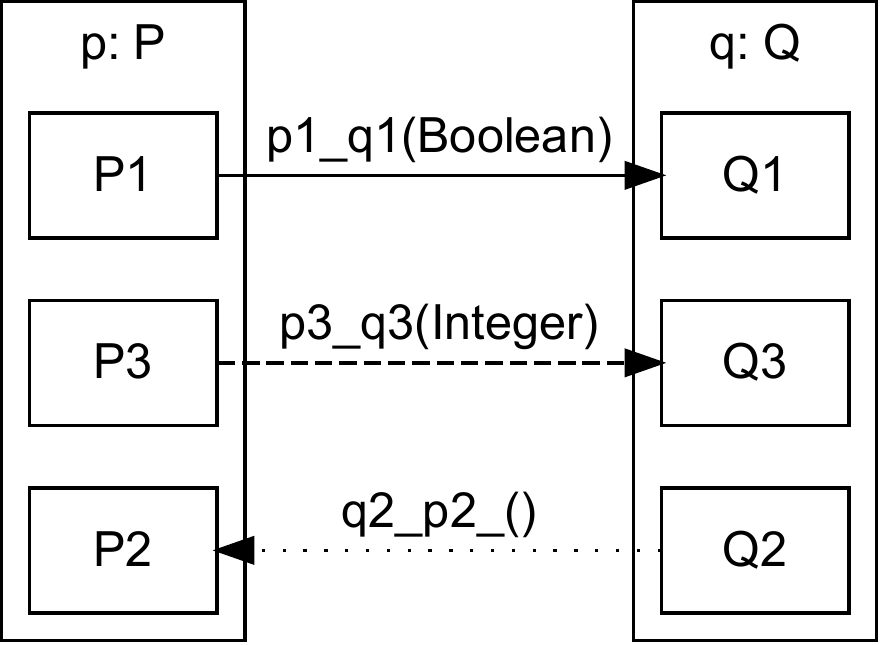}
\caption{Objects, ports and channels in $\SLCO$}
\label{fig:slco_example_communication}
  \end{minipage}
  \hspace{0.5cm}
  \begin{minipage}[b]{.45\textwidth}
    \centering
\includegraphics[width=1.1\columnwidth]{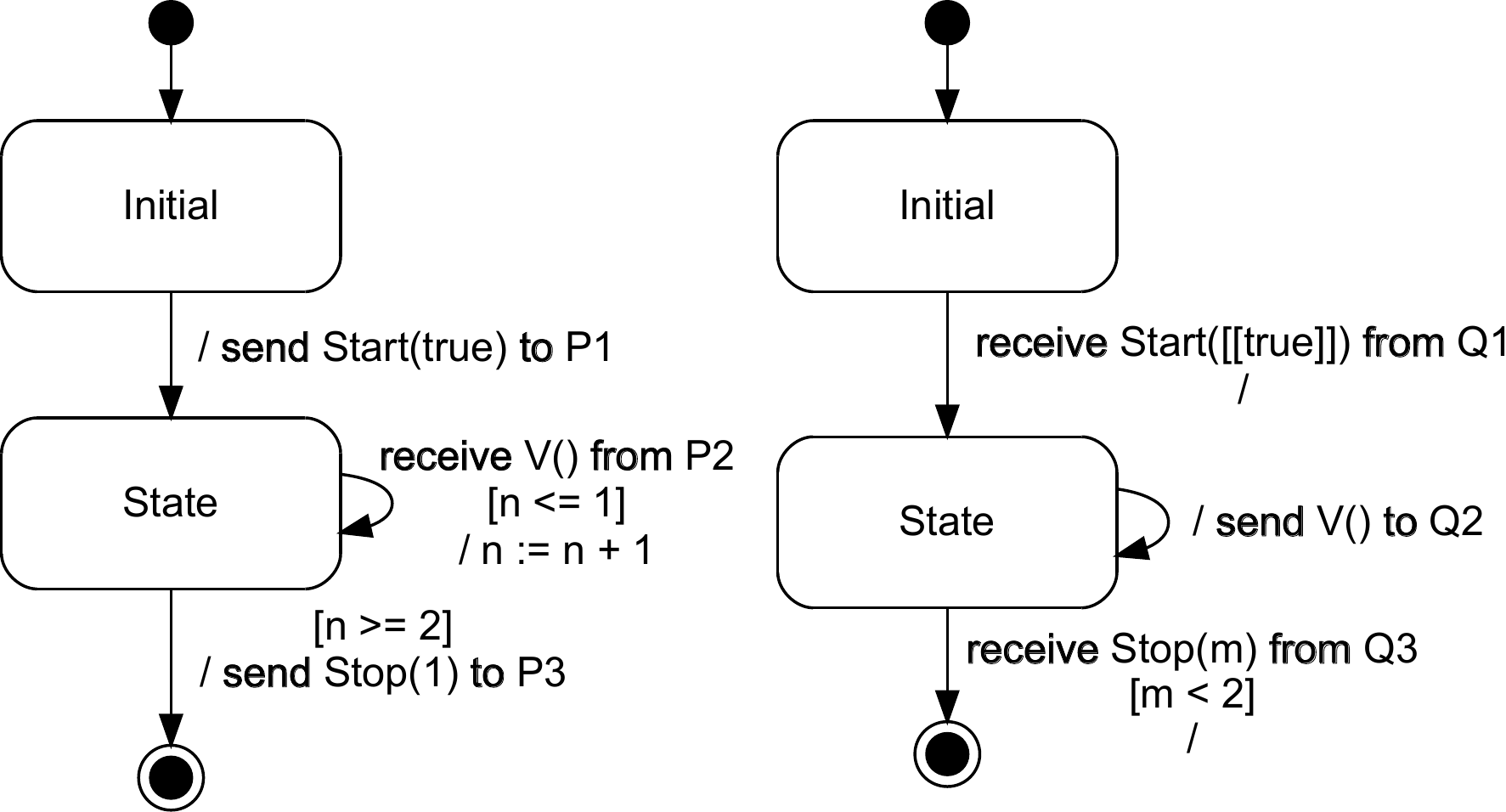}
\caption{Two $\SLCO$ state machines}
\label{fig:slco_example_sms}
  \end{minipage}
\end{figure}

An $\SLCO$ model consists of a number of classes, objects, and channels.
Objects are instances of classes and communicate with each other via channels, which are either bidirectional of unidirectional.
$\SLCO$ offers three types of channels: synchronous channels, asynchronous lossy channels, and asynchronous lossless channels.
An example of two objects connected by three channels is shown in Figure~\ref{fig:slco_example_communication}.
The objects~$\op$ and~$\oq$, which are instances of classes $P$ and~$Q$, can communicate over channels~$\op1\_\oq1$, $\oq2\_\op2$, and~$\op3\_\oq3$.
The arrows at the ends of the channels denote the direction of communication.
Synchronous channels are denoted by plain lines (e.g. $\op1\_\oq1$), asynchronous lossless channels are denoted by dashed lines (e.g. $\op3\_\oq3$), and asynchronous lossy channels are denoted by dotted lines (e.g. $\oq2\_\op2$).
A channel can only be used to send and receive signals with a certain signature, indicated by a number of argument types augmented to the channel.
Channel~$\op1\_\oq1$, for instance, can only be used to send and receive signals with a boolean argument, and channel~$\oq2\_\op2$ only allows signals without any arguments.

A class describes the structure and behavior of its instances.
A class has ports and variables that define the structure of its instances, and state machines that describe their behavior.
It is possible to specify the initial values of variables.
If no initial value is specified, integer variables are initialized to $0$, boolean variables are initialized to \emph{true}, and string variables are initialized to the empty string.
Ports are used to connect channels to objects.
Figure~\ref{fig:slco_example_communication} shows that object~$\op$ has ports $\it{P1}$, $\it{P2}$, and~$\it{P3}$, connecting it to channels~$\op1\_\oq1$, $\oq2\_\op2$, and~$\op3\_\oq3$, and that object $\oq$ has ports $\it{Q1}$, $\it{Q2}$, and~$\it{Q3}$, connecting it to the same channels.

A state machine consists of variables, states, and transitions.
$\SLCO$ allows for two special types of states: a state can be an initial state or a final state.
Each state machine has exactly one initial state, and can contain any number of ordinary and final states.
Figure~\ref{fig:slco_example_sms} shows an example of an $\SLCO$ model consisting of two state machines, whose initial states, $\Initial$, are indicated by a black dot-and-arrow, and whose final states are denoted as circled black dots.
As explained below, the left state machine specifies the behavior of object~$\op$ and the right state machine specifies the behavior of object~$\oq$, both already introduced  in~Figure~\ref{fig:slco_example_communication}.

A transition has a source and a target state, and possibly a guard, a trigger, a number of statements that form its effect, or a combination of these.
A guard is a boolean expression that must hold to enable the transition from the source to the target state to be taken.
For instance, $\texttt{[n >= 2]}$ is the guard of the transition with the source $\State$ and the final state as the target state in the state machine of $\op$.
There are two types of triggers: a signal reception and a delay.
A transition with a delay trigger is enabled after a specified amount of time has passed since entering its source state.
Note that our running example does not have this type of triggers.
A transition with a signal reception trigger is enabled if a signal is received via the port indicated by the trigger.
When a signal reception trigger is combined with a guard, naturally, the guard must hold for the transition to be enabled.
It is allowed for the guard to refer to arguments of the signal just being received, which yields a form of conditional message reception.
Take for instance the transition in $\oq$ from $\State$ to the final state, with trigger $\rec\ \Stop(m)\ \fromport\ Q3$.
It is only taken if the value of the argument sent with the signal $\Stop$ is smaller than 2, specified as $[m \texttt{<} 2]$.
Additionally, another form of conditional signal reception is offered.
Expressions given as arguments of a signal reception specify that only signals whose argument values are equal to the corresponding expressions are accepted.
Thus, $\oq$ in state $\Initial$  accepts only signals whose argument equals \emph{true}.

When a transition is made from one state to another state, the statements that constitute the effect of the transition are executed.
$\SLCO$ offers statements for assigning values to variables and for sending signals over channels.
The state machines in Figure~\ref{fig:slco_example_communication} specify the  following communication between $\op$ and $\oq$.
The two objects first communicate synchronously over channel~$\op1\_\oq1$, after which $\oq$ repeatedly sends signals to $\op$ over the lossy channel~$\oq2\_\op2$. As soon as $\op$ receives $2$ of the signals sent by $\oq$, it sends a signal over channel~$\op3\_\oq3$ and terminates.
After receiving this signal, $\oq$ terminates as well.


In addition to the graphical concrete syntax shown above, $\SLCO$ has a textual concrete syntax, which is used by the tools described in Section~\ref{sec:prototyping_semantics} to perform transformations of $\SLCO$ models.
Listing~\ref{lst:textual_model} shows a part of the textual equivalent of the model described above.

\begin{listing}[hbt]
\begin{centering}
\lstdefinelanguage{slco}{
  morekeywords={
    model, classes, variables, Integer, ports, machines, initial, state, final, transitions, from, to, effect, send,
    objects, channels, trigger, guard
  }
}

\lstset{
    language=slco,
    basicstyle=\scriptsize\ttfamily,
    keywordstyle=\bfseries,
    caption=Part of a textual $\SLCO$ model,
    captionpos=b,
    label=lst:textual_model
}
\begin{lstlisting}
model M {
  classes
    P {
      variables Integer n
      ports P1 P2 P3
      state machines
        P {
          initial Initial state State final Final
          transitions
            Receive from State to State {
              trigger receive V() from P2
              guard n <= 1
              effect n := n + 1 }
            ...
        }
    }
  ...
  objects p:P q:Q
  channels p1_q1(Boolean) sync from p.P1 to q.Q1
           ...
}
\end{lstlisting}
\end{centering}
\end{listing}

\vspace{-0.1cm}

%% file: Prototyping_Semantics1.tex
\section{Prototyping Semantics}
\label{sec:prototyping_semantics}
Figure~\ref{fig:languages_overview} shows that the $\SLCO$ transformation environment involves other languages and transformation tools.
Designing the environment and defining its main ingredients required, among others, thorough understanding of the $\SLCO$ semantics, and its specification in terms of basic activities, each one either being executed by a particular object or being the result of objects' interaction.
The idea behind the intermediate language $\CS$ is to specify explicitly these low-level activities, which are implicit in $\SLCO$, and to serve as an underlying language to express the semantics of $\SLCO$.
As a result, $\CS$ together with the $\SLCO 2 \CS$ transformation captures the semantics of $\SLCO$.
All languages and transformations described in this section are available for download~\cite{SLCOgooglecode}.

\subsection{Languages}
\label{sec:languages}
The process of transforming an $\SLCO$ model into an LTS is split into two steps.
First, the $\SLCO$ model is translated into a list of configurations and steps, represented in the $\CS$ language, where $\CSs$ stands for \emph{C}onfigurations and \emph{S}teps.
In this language, we describe the behavior of the entire system, resulting from the communication of the constituting objects, which are originally modeled as a set of separate state machines in $\SLCO$.
Then, this $\CS$ representation is transformed into a list of states and transitions, which form the $\LTS$ representation of the input model.

\begin{listing}[hbt]
\begin{centering}
\lstdefinelanguage{cs}{
 morekeywords={initial}
}

\lstset{
    language=cs,
    basicstyle=\scriptsize\ttfamily,
    keywordstyle=\bfseries,
    caption=The initial configuration of the running example,
    captionpos=b,
    label=lst:configuration
}
\begin{lstlisting}
<
 <p, P, Initial> <q, Q, Initial>,
 [<<p, n>,0>, <<q, m>,0>],
 [<<p3_q3, p, P3, q, Q3>,>, <<q2_p2, q, Q2, p, P2>,>],
 initial
>
\end{lstlisting}
\end{centering}
\end{listing}

\vspace{-0.8cm}

\paragraph{CS} The main ingredients in a $\CSs$ description are \emph{configurations} and \emph{steps}.
A configuration is a representation of a possible state of the system described by the $\SLCO$ model.
A configuration can make a step, after which the system reaches another configuration.
It is important to note that a step in $\CSs$ does not correspond to a transition in $\SLCO$, but, in general, several steps (not necessarily executed in sequence) accomplish the behavior specified with such an $\SLCO$ transition.
Therefore, the $\CSs$ representation of an $\SLCO$ model refines its behavior by splitting the execution in more basic activities.

Each configuration consists of three mandatory parts and an optional status.
The configuration given in Listing~\ref{lst:configuration} is the initial configuration of the model consisting of objects $\op$ and $\oq$, shown in Figures~\ref{fig:slco_example_communication} and~\ref{fig:slco_example_sms}.
The first part of a configuration specifies the current states of all state machines of all objects in the $\SLCO$ model.
This part of the configuration is referred to as the \emph{active states} of a configuration.
If a state machine \emph{sm} of an object \emph{o} is currently in state \emph{st}, this is specified as \hbox{\texttt{<o, sm, st>}} in the active states part of the configuration.
This type of active state is referred to as \emph{plain active state}.
Recall now that the effect of an $\SLCO$ transition may consist of several statements.
Thus, in general, the state machine can start executing the transition by receiving a signal, for instance, after which it starts executing the statements forming the effect of this transition.
All these parts of the transition's execution are represented as separate steps in a $\CSs$ model.
In view of the whole system, this means that the state machine to which this transition belongs to cannot proceed with any other transition as long as all statements are not executed; however, any other state machine is allowed to proceed with its own execution.
In order to specify these intermediate configurations in the $\CSs$ model, we introduce a second type of, so-called, \emph{partial active states}.
A partial active state \texttt{<o, sm, st, k, m>} represents that the \texttt{k}th statement in the list of statements is to be executed next as a part of the execution of the \texttt{m}th outgoing transition of state \emph{st}.
Consequently, the active states part of a configuration consists of a number of plain and partial active states, as many as the state machines in the model.

In the configuration shown in Listing~\ref{lst:configuration}, both active states, viz. \texttt{<p, P, Initial>} and \texttt{<q, Q, Initial>}, are plain.
Listing~\ref{lst:configuration_plain_partial} gives an example of a configuration in which \texttt{<q, Q, State>} is a plain active state and \hbox{\texttt{<p, P, State, 0, 1>}} is a partial active state, the latter indicating that the state machine of $\op$~(Figure~\ref{fig:slco_example_sms}) has to execute the 0th statement, \texttt{n:=n+1}, in the list of statements of the outgoing transition of state $\State$ whose identifier equals 1.

\begin{listing}[hbt]
\begin{centering}
\lstdefinelanguage{cs}{
 morekeywords={}
}

\lstset{
    language=cs,
    basicstyle=\scriptsize\ttfamily,
    keywordstyle=\bfseries,
    caption=A configuration containing a plain active state and a partial active state,
    captionpos=b,
    label=lst:configuration_plain_partial
}
\begin{lstlisting}
<
 <p, P, State, 0, 1> <q, Q, State>,
 [<<p, n>,0>, <<q, m>,0>],
 [<<p3_q3, p, P3, q, Q3>,>,<<q2_p2, q, Q2, p, P2>,>]
>
\end{lstlisting}
\end{centering}
\end{listing}

\vspace{-0.2cm}

The second part of the configuration, the \emph{valuation} part, maps variables to values.
In the example configuration in Listing~\ref{lst:configuration}, it is specified as \texttt{[<<p,n>,0>, <<q,m>,0>]}, expressing that variable \emph{n} of object $\op$ and variable \emph{m} of object $\oq$ both have value 0 in this configuration.

The third part of the configuration represents a set of buffers and is simply referred to as the \emph{buffers} of the configuration.
For each asynchronous channel in the model, one or two buffers are introduced:~in case of a bidirectional channel two buffers are introduced and in case of a unidirectional channel one buffer is introduced.
The configuration in Listing~\ref{lst:configuration} contains two buffers, one for each (unidirectional) channel, and they are both empty.
The first buffer corresponds to channel~$\op3\_\oq3$ and the second buffer corresponds to channel~$\oq2\_\op2$.
Because channel~$\op1\_\oq1$ is synchronous, it has no corresponding buffer.
The optional status of this configuration is set to \emph{initial}.
If all state machines in a configuration are in a final state, the status of this configuration is set to \emph{final}.

The dynamics of the system modeled by a set of communicating objects is represented by \emph{steps}.
Each step has a source and a target configuration, and an optional label.
A step is a representation of a (basic) activity a given state machine has to perform as a part of the set of activities assigned to a transition in the $\SLCO$ representation of this state machine.
Listing~\ref{lst:steps} shows two steps from the $\CSs$ model of our $\SLCO$ running example model.
The first step has a label that represents the reception of the asynchronous signal~$V$ by the state machine of $\op$.
The second step represents the execution of the assignment statement \texttt{n := n + 1} of this state machine and does not have a label.

\begin{listing}[hbt]
\begin{centering}
\lstdefinelanguage{cs}{
 morekeywords={initial}
}

\lstset{
    language=cs,
    basicstyle=\scriptsize\ttfamily,
    keywordstyle=\bfseries,
    caption=Two steps depicting the reception of a signal and the execution of an assignment statement,
    captionpos=b,
    label=lst:steps
}
\begin{lstlisting}
<
 <
  <p, P, State> <q, Q, State>, [<<p, n>,0>, <<q, m>,0>],
  [<<p3_q3, p, P3, q, Q3>,>, <<q2_p2, q, Q2, p, P2>,<V, >>]
 >,
 "receiving V()",
 <
  <p, P, State, 0, 1> <q, Q, State>, [<<p, n>,0>, <<q, m>,0>],
  [<<p3_q3, p, P3, q, Q3>,>, <<q2_p2, q, Q2, p, P2>,>]
 >
>
<
 <
  <p, P, State, 0, 1> <q, Q, State>, [<<p, n>,0>, <<q, m>,0>],
  [<<p3_q3, p, P3, q, Q3>,>,<<q2_p2, q, Q2, p, P2>,>]
 >,
 <
  <p, P, State> <q, Q, State>, [<<p, n>,1>, <<q, m>,0>],
  [<<p3_q3, p, P3, q, Q3>,>,<<q2_p2, q, Q2, p, P2>,>]
 >
>
\end{lstlisting}
\end{centering}
\end{listing}

\paragraph{LTS}
The language $\LTS$ is a simple language for representing labeled transition systems as a list of states and a list of transitions.
A state can be typed as initial or final state.
Each transition is described as a couple of states, the source and the target states, and an optional label.
Listing~\ref{lst:lts} shows the $\LTS$ description of a tiny LTS  with four states and three transitions.
State $0$ is declared as an initial state, and states $1$ and $3$ are final states.
There are three transitions, one of which has label $a$.
There are different languages that describe LTSs; the ones used in our environment are described in Section~\ref{sec:verification_and_visualization}.
The most notable feature of $\LTS$ in comparison to these other languages is that it has a notion of final state, to distinguish between successful termination and deadlock.
A state without any outgoing transitions represents a deadlock.
It is considered undesirable for a system to reach these deadlock states.
Reaching a final state and thus terminating successfully, however, is considered desirable behavior.

\begin{listing}[hbt]
\begin{centering}
\lstdefinelanguage{lts}{
 morekeywords={initial, states, final, transitions}
}

\lstset{
    language=lts,
    basicstyle=\scriptsize\ttfamily,
    keywordstyle=\bfseries,
    caption=An example of an LTS,
    captionpos=b,
    label=lst:lts
}
\begin{lstlisting}
states
  initial 0
  final 1
  2
  final 3
transitions
  0 1
  0 "a" 2
  2 3
\end{lstlisting}
\end{centering}
\end{listing}

\vspace{-0.5cm}

\subsection{Tools}
Now that the languages used to prototype $\SLCO$ are in place, we describe the two main transformations of our environment in this subsection.
The first transformation produces a $\CSs$ representation of an input $\SLCO$ model and the second transformation produces an LTS from the $\CSs$ representation (Figure~\ref{fig:languages_overview}).
As these transformations are implemented in the ASF+SDF Meta-Environment~\cite{Brand:2001:ASF}, we briefly describe it here as well.

\paragraph{ASF+SDF}
\label{par:asf+sdf}
The language ASF+SDF is a combination of the two formalisms ASF~\cite{AlgebraicSpecification1989} and SDF~\cite{Visser97syntaxdefinition}.
SDF stands for \emph{Syntax Definition Formalism}.
It is a formalism for defining syntax of context-free languages.
ASF stands for \emph{Algebraic Specification Formalism}.
It is a formalism for the definition of conditional rewrite rules.
Given a syntax definition in SDF of the source and target language, ASF can be used to define a transformation from the source language to the target language.
In ASF, conditional rewrite rules are specified using the concrete syntax of the input and output languages.

ASF in combination with SDF guarantees syntax safety of the implementation of the transformations.
A transformation is syntax safe if it only accepts input that adheres to the syntax definition of the input language, and it always produces output that adheres to the syntax of the output language.

The ASF+SDF Meta-Environment is an IDE for ASF+SDF.
It has a graphical user interface that offers syntax-highlighting for the specification of SDF and ASF definitions, and an interpreter and debugger for the execution and debugging of ASF specifications.
The Meta-Environment can be used to create a command-line tool that parses and rewrites input adhering to the syntax definition of the input language, and outputs the result.
These command-line tools employ memoization, which ensures that the result of a rewrite rule applied to a given term is computed only once.
Both the ASF+SDF Meta-Environment and the command-line tools it generates use \emph{Annotated Terms} (ATerms)~\cite{Brand:2007:AME:1219180.1219600} to represent terms internally.
Because ATerms offer maximal subterm sharing, the internal representation of terms uses as little space as possible.

\paragraph{SLCO2CS}
An $\SLCO$ model is translated into $\CSs$ in three phases.
First, the initial configuration of the $\SLCO$ model is constructed.
The list of active states of this initial configuration consists of the initial states of each of the state machines of the objects in the model, the valuation maps all variables to their initial values, and the buffers corresponding to all asynchronous channels are empty.
Second, the set of all reachable configurations is generated.
This phase is described in more detail below.
Third, the list of configurations is traversed to find the configurations containing only active states that are final, and these configurations are marked as final.

In the second phase, first all configurations that are reachable from the initial configuration are created, as well as all the steps from the initial configuration to the reachable configurations.
Then, all configurations that are reachable from these new configurations and the corresponding steps are created, and so on, until no new configuration is found.
The configurations that are reachable from a source configuration are computed based on the active states of this source configuration.
A step from one configuration to another is possible if one of the active states of the source configuration has an outgoing transition that is enabled or can execute a statement, in the corresponding $\SLCO$ state machine.
Whether a transition is enabled depends on the valuation of the variables and the contents of the buffers of the source configuration.
The valuation of the variables is used to determine whether the optional guard of such a transition holds and the content of the buffers is used to determine whether any signal receptions are possible.
To determine all the configurations that are reachable from a given source configuration, all outgoing transitions of all active states of the source configuration are considered.
The SDF functions discussed next are selected from the set of all functions that all together implement the generation of configurations and steps within the second phase of the transformation.

Listing~\ref{lst:sdf_steps} shows the signature of the functions \emph{takeStepTransition} and \emph{takeStepPartialTransition}.
Applying these functions to terms representing a model, a configuration, an active step, and a transition results in a term representing a list of configurations and a list of steps.
The function \emph{takeStepTransition} is meant for handling plain active states, whereas the function \emph{takeStepPartialTransition} is meant for handling partial active states.

\begin{listing}[hbt]
\begin{centering}
\lstdefinelanguage{sdf}{
 morekeywords={}
}

\lstset{
    language=sdf,
    basicstyle=\scriptsize\ttfamily,
    keywordstyle=\bfseries,
    caption=SDF definition of the functions that compute all possible steps from configurations,
    captionpos=b,
    label=lst:sdf_steps
}
\begin{lstlisting}
takeStepTransition(
  Model, Configuration, ActiveState, Transition
) -> <Configuration*, Step*>

takeStepPartialTransition(
  Model, Configuration, ActiveState, Transition
) -> <Configuration*, Step*>
\end{lstlisting}
\end{centering}
\end{listing}

Listing~\ref{lst:asf_signalreception} shows one of the conditional rewrite rules in ASF that implement the function \emph{takeStepTransition}.
In the rules in Listing~\ref{lst:asf_signalreception}, \ref{lst:asf_assignment}, and~\ref{lst:asf_partial}, all variable names start with a dollar sign.
In ASF+SDF, each term conforms to a \emph{sort} and variables represent arbitrary terms of some sort.
Variable names that end with a plus symbol represent lists of more than one terms of a sort and variable names that end with a question mark represent zero or one term.
For example, the variables \texttt{\$Statement+} and \texttt{\$Guard?} in Listing~\ref{lst:asf_signalreception} represent one or more statements and an optional guard, respectively.
The first part of an ASF rule consists of the conditions of that rule.
After an arrow (\texttt{====>}), the left-hand side and right-hand side of the rule follow, separated by an equal sign.
If all the conditions hold, the left-hand side can be replaced by the right-hand side.
The rules described in this paper use only one kind of condition: a matching condition.
A matching condition consists of a right-hand side and a left-hand side, separated by a colon and an equal sign.
The condition holds if both sides can be matched.
In this case, the variables occurring at the right-hand side are instantiated such that both sides match.

The rule in Listing~\ref{lst:asf_signalreception} deals with transitions with a signal reception trigger and an effect consisting of more than one statement. The source state of such a transition is represented by a plain active state.
Because the effect of the transition consists of several statements, the plain active state provided as input is replaced by a partial active state in the resulting configuration.
This partial active state indicates that none of the statements that form the effect have been executed yet and that the transition that has been taken from state \emph{\$ActiveState} has identifier \emph{\$NatCon}.
The first step in Listing~\ref{lst:steps} is produced by applying this rule.

\begin{listing}[hbt]
\begin{centering}
\lstdefinelanguage{asf}{
 morekeywords={from, to, trigger, effect}
}

\lstset{
    language=asf,
    basicstyle=\scriptsize\ttfamily,
    keywordstyle=\bfseries,
    caption=ASF rule that specifies how a signal reception trigger on a transition from a plain active state is processed,
    captionpos=b,
    label=lst:asf_signalreception
}
\begin{lstlisting}
[takeStepTransition-reception-1]
  <$IdCon0, $IdCon1, $IdCon2> := $ActiveState,
  $IdCon3 from $IdCon2 to $IdCon4 {
    trigger $SignalReception
    $Guard?
    effect $Statement+
  } := $Transition,
  $NatCon := getTransitionNumber($Model, $Transition, $ActiveState),
  $ActiveState0 := <$IdCon0, $IdCon1, $IdCon2, 0, $NatCon>,
  <$Configuration0, $Step0> := processSignalReception(
    $SignalReception, $ActiveState, $ActiveState0, $Model, $Configuration,
    $IdCon0, $IdCon1
  )
  ====>
  takeStepTransition(
    $Model, $Configuration, $ActiveState, $Transition
  ) = <$Configuration0, $Step0>
\end{lstlisting}
\end{centering}
\end{listing}

Listing~\ref{lst:asf_assignment} shows yet another of the conditional rewrite rules that implements the function \emph{takeStepTransition}.
This rule applies to transitions that have no trigger and only one single assignment statement that forms its effect.
In the configuration resulting from this rule, the original active state is replaced by another plain active state, because the effect consists of only one statement.
The function \emph{processAssignmentStatement}, used by the function \emph{takeStepTransition}, produces a configuration and a step.
The configuration is an updated version of the configuration provided as input, in which the active state \emph{\$ActiveState} is replaced by \emph{\$ActiveState0} and the valuation is adapted according to the assignment statement.
The step consists of the original configuration and the updated configuration.

\begin{listing}[hbt]
\begin{centering}
\lstdefinelanguage{asf}{
 morekeywords={from, to, effect}
}

\lstset{
    language=asf,
    basicstyle=\scriptsize\ttfamily,
    keywordstyle=\bfseries,
    caption=ASF rule that specifies how a single assignment statement on a transition from a plain active state is processed,
    captionpos=b,
    label=lst:asf_assignment
}
\begin{lstlisting}
[takeStepTransition-assignment-0]
  <$IdCon0, $IdCon1, $IdCon2> := $ActiveState,
  $IdCon3 from $IdCon2 to $IdCon4 {
    $Guard?
    effect $AssignmentStatement
  } := $Transition,
  $ActiveState0 := <$IdCon0, $IdCon1, $IdCon4>,
  <$Configuration0, $Step0> := processAssignmentStatement(
     $AssignmentStatement, $ActiveState, $ActiveState0, $Configuration,
     $IdCon0, $IdCon1
  )
  ====>
  takeStepTransition(
    $Model, $Configuration, $ActiveState, $Transition
  ) = <$Configuration0, $Step0>
\end{lstlisting}
\end{centering}
\end{listing}

Listing~\ref{lst:asf_partial} deals with transitions with an optional guard, an optional trigger, and an effect consisting of more than one statement.
The source state of this transition is a partial active state.
The function \emph{getStatements} that is used by this rule takes a list of statements and an transition identifier and returns the statement with the given index and all following statements.
One of the matching conditions states that this rule only applies if the statement with index \emph{\$NatCon0} is an assignment statement and that there are no statements after this statement.
The second step of Listing~\ref{lst:steps} is the result of applying this rule to the target configuration of the first step.

\begin{listing}[hbt]
\begin{centering}
\lstdefinelanguage{asf}{
 morekeywords={from, to, effect}
}

\lstset{
    language=asf,
    basicstyle=\scriptsize\ttfamily,
    keywordstyle=\bfseries,
    caption=ASF rule that specifies how an assignment statement on a transition from a partial active state is processed,
    captionpos=b,
    label=lst:asf_partial
}
\begin{lstlisting}
[takeStepsPartialTransition-assign-0]
  $IdCon3 from $IdCon2 to $IdCon4 {
    $Trigger?
    $Guard?
    effect $Statement+
  } := $Transition,
  $AssignmentStatement := getStatements($Statement+, $NatCon0),
  $ActiveState0 := <$IdCon0, $IdCon1, $IdCon4>,
  <$Configuration0, $Step0> := processAssignmentStatement(
    $AssignmentStatement, <$IdCon0, $IdCon1, $IdCon2, $NatCon0, $NatCon1>,
    $ActiveState0, $Configuration, $IdCon0, $IdCon1
  )
  ====>
  takeStepPartialTransition(
    $Model, $Configuration, <$IdCon0, $IdCon1, $IdCon2, $NatCon0, $NatCon1>, $Transition
  ) = <$Configuration0, $Step0>
\end{lstlisting}
\end{centering}
\end{listing}

\paragraph{CS2LTS}
The tool $\CS 2\LTS$ translates lists of configurations and steps from $\CSs$ to $\LTS$, as shown in Figure~\ref{fig:languages_overview}, and it is also implemented using ASF+SDF.
The way we have defined $\CSs$ makes this translation rather straightforward.
Each configuration is mapped to a unique natural number and, possibly, an optional status.
The optional status indicates whether a state in the LTS is an initial or a final state and it is equal to the status of the configuration that corresponds to the state.
Each step is transformed to a pair of natural numbers representing its configurations, possibly decorated by an optional label.
The label of a transition is equal to the label of the corresponding step.

%% file: Verification.tex
\section{Verification and Visualization}
\label{sec:verification_and_visualization}

\subsection{Visualization}

\begin{figure}[hbt]
  \begin{minipage}[b]{.45\textwidth}
    \centering
    \includegraphics[width=.6\textwidth]{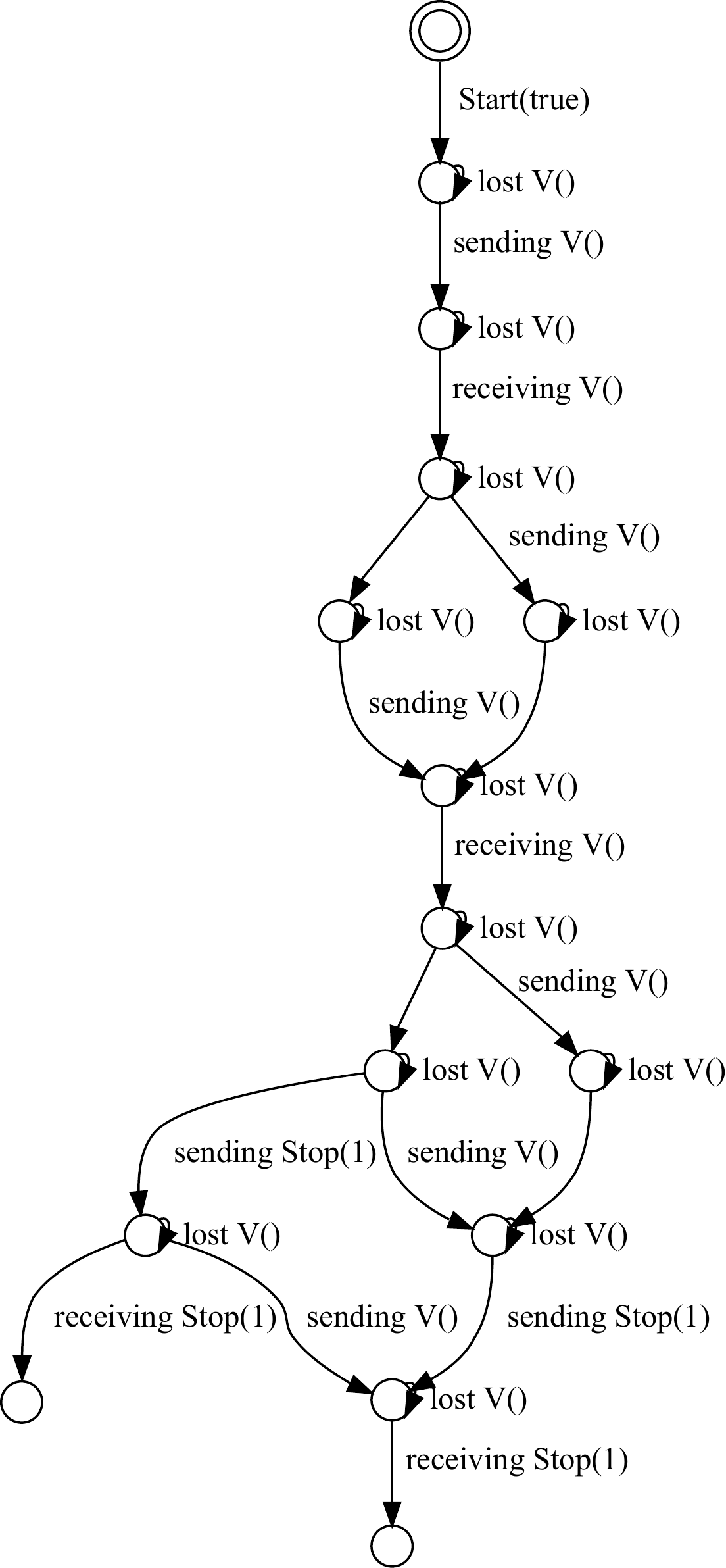}
    \caption{An LTS visualized using $\DOT$}
    \label{fig:statespace}
  \end{minipage}
  \hspace{0.5cm}
  \begin{minipage}[b]{.45\textwidth}
    \centering
    \includegraphics[width=.2\textwidth]{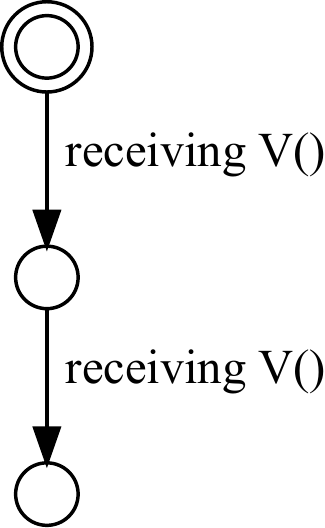}
    \caption{A reduced LTS visualized using $\DOT$}
    \label{fig:statespace_reduced}
  \end{minipage}
\end{figure}

The $\DOT$ language is a language for graph visualization that is part of the graphviz toolset~\cite{Ellson01graphviz—}.
We use this language to visualize LTSs, by translating LTSs to graphs in the format of the $\DOT$ language.
This translation is straightforward, because the way graphs are described in this language is similar to the description of LTSs.
A graph description in $\DOT$ is a list of nodes and edges from node to node, combined with attributes that specify how particular nodes and edges should be displayed.
These attributes define, for example, the color, width, height, and the type of line used to draw these nodes and edges.
LTSs are transformed to $\DOT$ graphs by translating all states to nodes and all transitions to edges, and specifying how initial states, final states, and normal states should be visualized.
Figure~\ref{fig:statespace} shows the LTS that corresponds to the model in Figure~\ref{fig:slco_example_communication} and~\ref{fig:slco_example_sms}, visualized using $\DOT$.
In the visual representation, a transition label is always place on the right-hand side of the transition.
Several transitions do not have labels, as they correspond to assignment statements.

\subsection{Verification}
For small LTSs, like the one in Figure~\ref{fig:statespace}, it is easy to verify properties manually by inspecting the graph.
In case of larger LTSs, reduction techniques can be applied to first reduce the LTS before verifying properties manually.
The tool \emph{ltsconvert}, which is part of the mCRL2 toolset~\cite{DBLP:conf/dagstuhl/GrooteMRUW06}, takes an LTS in various formats as input and converts it to an equivalent LTS in another format.
One of the formats that \emph{ltsconvert} is able to process is the $\DOT$ format, 
\mcc{which although meant to describe graphical representations of graphs, is also used to represent LTSs.}
The tool is also capable of reducing LTSs, by means of an equivalence relation, and several equivalence relations are supported.
Figure~\ref{fig:statespace_reduced} shows the LTS obtained after reduction has been applied to the LTS in Figure~\ref{fig:statespace}, by means of the branching bisimilarity relation~\cite{GW96}.
The LTS has been reduced by, first, turning all labels except \texttt{receiving~V()} to internal unobservable labels and then removing all redundant states and transitions using \emph{ltsconvert}.
A similar procedure can be applied for other reductions this tool supports.
\mcc{After the reduction, it is obvious that the following property holds in our running example: ``signal $V$ is received exactly two times.''}

As mentioned in Section~\ref{sec:Introduction}, there exists a number of transformations that transform $\SLCO$ models into other $\SLCO$ models with equivalent observable behavior.
By producing an LTS for the input and the output model of such a transformation, and reducing both LTSs using the technique described above, the correctness of this transformation for the given input model can be verified by comparing the reduced LTSs.
If reducing the LTS of both models leads to the same LTS, the transformation has indeed preserved the observable behavior.

When LTSs get too large for reduction and manual inspection, other tools can be used for verification.
One approach is converting LTSs to the \emph{BCG} and \emph{AUT} file formats that are used by the CADP toolset~\cite{DBLP:conf/tacas/GaravelLMS11} to represent LTSs.
The CADP toolset offers tools that take an LTS and a temporal logic property as input and perform on-the-fly verification of the property on the LTS. Alternatively, the previously mentioned mCRL2 toolset can be used for verification too.
This toolset includes a tool that can transform LTSs to the proprietary format of the toolset, and tools that can be used to analyze, simulate, manipulate, and visualize models described using this format.
These two example toolsets clearly show the added benefit of producing LTSs from $\SLCO$ models.
Transforming models to this common description format makes it possible to verify properties of models using existing tools, without additional effort. 

%% file: Related_Work.tex
\section{Related Work}
\label{sec:Related_Work}
Hooman and Van der Zwaag~\cite{Hooman2006} used the interactive theorem prover PVS to define the semantics of a subset of the UML.
In this subset, the behavior of objects is specified using state machines that communicate with each other both synchronously and asynchronously.
Proving properties of models in this approach is done manually using PVS.
This is a complex task that requires expertise in PVS, which can be simplified using certain predefined strategies.
A disadvantage of this approach is that is does not offer the reusability of other existing tools that our approach offers.
An advantage of this approach is that it does not suffer from the state-space explosion problem, because the complete state space of models does not have to be generated for property verification.

Di Ruscio~\cite{Ruscio06apractical} et al.\ define the semantics of a DSL for the development of telephony services using Abstract State Machines (ASM).
Because ASMs can be executed, this definition can be used to simulate models specified in their DSL.
The approach is meant for the specification of the behavioral semantics of the DSL only, and does not offer verification of models.
Proving properties for all models in general or any specific model is not investigated.
In theory, however, properties of models could be verified in the domain of ASMs.

Sadilek and Wachsmuth~\cite{Sadilek:2008:PVI:1426334.1426341} propose a technique for defining the semantics of DSLs that uses model instances as configurations and QVT relations to define steps between configurations.
Configurations, representing model instances, can be visualized using the same editors used to create models.
By reusing the existing editors, visual interpreters and visual debuggers can be created with relatively little effort.
Although this technique is suited for simulation of models, it is not efficient enough for state-space generation.
Because each configuration is represented by a model, a lot of memory is needed to store all possible configurations.

A number of approaches use Maude to specify the operational semantics of DSLs~\cite{Rusu:2011:EDM:1921532.1921557, Rivera:2009:FSA:1631662.1631666}.
Given the operational semantics of a DSL in Maude, other techniques can be applied to verify properties of DSL models.
Both an LTL model checker~\cite{wrla2002:mmc} and a $\mu$-calculus model checker~\cite{Wang:2005:9MC:1705545.1705992} are available for rewrite systems specified in Maude.
\mcc{Although it is clear that model checking techniques can be implemented in Maude and applied to specifications of the semantics of DSLs, not all techniques applicable to LTSs which we aim to exploit, such as reduction and visualization, have been implemented in Maude.}
It might be the case, therefore, that a given technique must first be implemented in Maude before it can be used in combination with a specification of the semantics of a DSL.
\mcc{With our approach, we can connect to various tools and apply existing techniques only by adapting the representation of LTSs, if needed.}

\vspace{-0.3cm}

%% file: Conclusions_and_Future_Work.tex
\section{Conclusions and Future Work}
\label{sec:Conclusions_and_Future_Work}
Defining the semantics of $\SLCO$ by implementing a transformation that transforms $\SLCO$ models to LTSs has a number of advantages.
Existing tools for verification and visualization can be reused, because LTSs are a common input representation used by a number of tools.
The biggest advantages of using the ASF+SDF Meta-Environment for this implementation are ATerms, used for representing terms, and the command-line tools that can be automatically generated.
Using ATerms guarantees efficient use of memory and the command-line tools offer efficient execution of rewrite rules, without any additional effort during the implementation.
Both execution speed and efficient use of memory are important in this case because the state spaces of models represented by LTSs are typically very large.

As future work we want to define a formal semantics of $\SLCO$ using, for instance, structural operational semantics.
The prototype of the semantics of $\SLCO$ and its implementation as presented here, make a solid ground for this work, as we got better understanding of the semantics of this DSL and were able to investigate a number of design decisions.
As $\SLCO$ has the notion of time delays, future work is also going to investigate possibilities to connect our $\SLCO$ environment to languages and tools for time analysis.
We also consider to apply the approach taken in this paper to other DSLs.
\mcc{The approach lends itself well for the creation of state-space generators based on the operational semantics of a given DSL, and prototyping the semantics of languages with informal or incompletely defined operational semantics.
}

\vspace{-0.7cm}